\documentclass{cargese}
\usepackage{graphicx}
\usepackage[english]{babel}

\let\footnote\savefootnote
\let\footnotetext\savefootnotetext
\let\footnoterule\savefootnoterule
\normallatexbib
\usepackage{amsmath, amssymb, amsfonts}
\usepackage{epsfig}

\begin{document}
%\numberwithin{equation}{section}
%\numberwithin{figure}{subsection}

\articletitle[Closed strings in Misner space]{ Closed strings in
 Misner space: \\ a toy model for a Big Bounce ? \footnote{To
appear in the Proceedings of the NATO ASI and EC Summer School
``String Theory: from Gauge Interactions to Cosmology'', Carg\`ese,
France, June 7-19, 2004. \hfill {\tt hep-th/0501145}}
} 

\chaptitlerunninghead{Closed strings in Misner space}

\author{Bruno Durin$^\dagger$ and Boris Pioline$^{\dagger\star}$}

\affil{$^\dagger$ LPTHE, Universit\'e Paris 6, \\
4 place Jussieu, 75252 Paris cedex 05, France}

\affil{$\star$ LPTENS, D\'epartement de 
Physique de l'Ecole Normale Sup\'erieure \\
24 rue Lhomond, 75231 Paris cedex 05, France}

\email{email: bdurin@lpthe.jussieu.fr, pioline@lpthe.jussieu.fr}

\abstract{Misner space, also known as the Lorentzian orbifold
$R^{1,1}/boost$, is one of the simplest examples of a cosmological 
singularity in string theory. In this lecture, we review the 
semi-classical propagation of closed strings in this background,
 with a particular emphasis on the twisted sectors of the orbifold.
 Tree-level scattering amplitudes and the one-loop vacuum amplitude 
 are also discussed.}

\normalsize\leftskip=0pt

\vskip .8cm
\begin{flushright}
{\it\scriptsize
Thus I was moving along the sloping curve of the time loop\\ towards
that place in which the Friday me before  the beating\\ would change into
the Friday me already beaten. \\ I. Tichy, \cite{lem}}
\end{flushright}

\vskip .4cm

Despite their remarkable success in explaining a
growing body of high precision cosmological data, 
inflationary models, just as the Hot Big Bang Model,
predict an Initial Singularity where effective field
theory ceases to be valid \cite{Borde:1993xh}. As a quantum theory of
gravity, String Theory ought to make sense 
even in this strongly curved regime, possibly by providing an initial 
quantum state if the Initial Singularity is truely an Origin of
Time, or by escaping it altogether if stringy
matter turns out to be less prone to gravitational collapse than 
conventional field-theoretic matter. Unfortunately, describing cosmological 
singularities and, less ambitiously, time dependence in string theory
has been a naggingly difficult task, partly because of the absence
of a tractable closed string field theory framework. Unless
stringy ($\alpha'$) corrections in the two-dimensional sigma model
are sufficient to eliminate the singularity, quantum ($g_s$)
corrections are expected to be important due to the large blue-shift
experienced by particles or strings as they approach the singularity, 
invalidating a perturbative approach. Nevertheless, one may expect cosmological
production of particles, strings and other extended states 
near the singularity to qualitatively alter the dynamics, and it is
not unplausible, though still speculative, that their contribution 
to the vacuum energy be sufficient to lead to a Big Bounce 
rather than an Big Bang.

In order to make progress on this issue, it is useful to study toy
models where at least $\alpha'$ corrections are under control, and
study string production to leading order in $g_s$. Orbifolds, being
locally flat, are immune to $\alpha'$ corrections, and thus a good
testing playground. One of the simplest examples of time-dependent 
orbifolds\footnote{The orbifold of $R^{1,1}$ under time reversal 
may be even simpler, but raises further puzzles related to time 
unorientability \cite{Balasubramanian:2002ry}. Discussions of other exact
cosmological backgrounds in string theory include
\cite{Antoniadis:1988aa,
Nappi:1992kv, Kounnas:1992wc, Kiritsis:1994fd, 
Elitzur:2002rt,Cornalba:2002fi,Craps:2002ii,
Dudas:2002dg,Cornalba:2003kd,Johnson:2004zq,Toumbas:2004fe}.}
is the Lorentzian 
orbifold $\mathbb{R}^{1,1}/boost$ \cite{Horowitz:ap,Khoury:2001bz,
Nekrasov:2002kf}, 
formerly known as Misner space \cite{Misner} in the gravity literature.
Introduced as a local model for the cosmological singularity 
and chronological horizon of Lorentzian Taub-NUT space, 
Misner space was shown long ago to
exhibit divergences from quantum vacuum fluctuations in field theory,
at least for generic choices of vacua \cite{Hiscock:vq}. 
Not surprisingly, this is also true in string theory, although less 
apparent since the local value of the energy-momentum tensor is not 
an on-shell observable \cite{Berkooz:2004re}. 
Similarly, just as in field theory, tree-level scattering amplitudes 
of field-theoretical (untwisted) states have been found to diverge, 
as a result of large graviton exchange near the 
singularity \cite{Berkooz:2002je}.

While these facts ominously indicate that quantum back-reaction 
may drastically change the character of the singularity, experience from 
Euclidean orbifolds suggests that twisted states may alleviate the 
singularities of the effective field theory description, and that it
may be worthwhile to investigate their classical behaviour, overpassing
the probable inconsistency of perturbation theory. Indeed, in the
context of Misner space, twisted states are just strings that wind around
the collapsing spatial direction, and become the lightest degrees of
freedom near the singularity. In these notes, we review classical aspects
of the propagation of closed strings in Misner space, with particular
emphasis on twisted states, based on the recent works
\cite{Pioline:2003bs, Berkooz:2004re, Berkooz:2004yy}.

The outline is as follows. In Section 2, we 
describe the semi-classical dynamics of charged particles and winding
strings, and compute their cosmological 
production rate, at tree level in the singular Misner geometry -- although 
our approach is applicable to more general cases. 
In Section 3, we analyze the imaginary part of the one-loop
amplitude, which carries the same information in principle.
In Section 4, we review recent results on scattering amplitudes 
of untwisted and twisted states, and their relation to 
the problem of classical back-reaction from a 
``condensate'' of twisted states. Section 5 contains our 
closing remarks.

\section{Semi-classics of closed strings in Misner space}

\subsection{Misner space as a Lorentzian orbifold}

\begin{figure}
\begin{center}
\epsfig{file=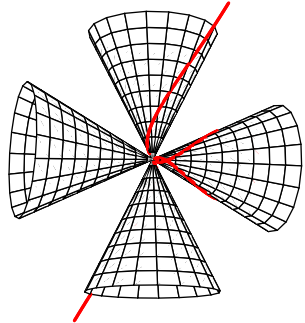,height=5cm}
\caption{\label{geomisner}
Free particles or untwisted strings propagate from the past Milne
region to the future Milne region, with a temporary excursion in the 
whiskers.}
\end{center}
\end{figure}

Misner space was first introduced in the gravity literature as a local model 
\cite{Misner} for the singularities of the Taub-NUT space-time \cite{Taub} . 
It can be formally defined as
the quotient of two-dimensional\footnote{Higher dimensional analogues
have also been considered \cite{Russo:2003ky}.}
Minkowski space $\mathbb{R}^{1,1}$ by the finite
boost transformation $B:\ (x^+,x^-)\rightarrow (e^{2\pi\beta} x^+, 
e^{-2\pi\beta}x^-)$, where $x^\pm$ are the light-cone coordinates. As such,
it is a locally flat space, with curvature localized at the fixed locus 
under the identification, i.e. on the light-cone $x^+ x^- = 0$.
The geometry of the quotient can be pictured as four Lorentzian cones touching 
at their apex (See Figure \ref{geomisner}), corresponding to the four quadrants of the covering space
$\mathbb{R}^{1,1}$. Choosing coordinates adapted to the boost $B$,
\begin{align}
x^\pm  &= T e^{\pm \beta
\theta}/\sqrt{2}\ , & & x^+ x^->0
\quad \mbox{(Milne regions)} \label{xpm1} \\
x^\pm  &= \pm r e^{\pm \beta \eta}/\sqrt{2}\
, & & x^+ x^-<0 \quad \mbox{(Rindler regions)} &\label{xpm2}
\end{align}
where, due to the boost identification, the coordinates $\theta$ and $\eta$
are compact with period $2\pi$, the metric of the quotient can be written as  
\begin{equation} 
ds^2 = -2\ dx^+ dx^-
 = \left\{
\begin{matrix}
-dT^2 + \beta^2 T^2 d\theta^2\\
dr^2 - \beta^2 r^2 d\eta^2
\end{matrix}
\right\}
\end{equation}
The two regions $x^+<0, x^-<0$ (P) and $x^+>0, x^->0$ (F),
describe contracting and expanding cosmologies where the radius of the spatial 
circle parameterized by $\theta$ changes linearly in time, and are often
called (compactified) Milne regions. The space-like cones $x^+>0, x^-<0$ (R) 
and $x^+<0, x^->0$ (L), 
often termed ``whiskers'', are instead time-independent 
Rindler geometries with compact time $\eta$ \footnote{This should not be
confused with thermal Rindler space, which is periodic in {\it imaginary} time.}. 
The Milne and Rindler regions, tensored with a sphere of finite size, 
describe the Taub and NUT regions, respectively, of the Taub-NUT space-time 
in the vicinity of one of its infinite sequence of cosmological
singularities. It also captures the local geometry in a variety of 
other cosmological string backgrounds
\cite{Nappi:1992kv,Elitzur:2002rt,
Craps:2002ii,Dudas:2002dg}.
It is also interesting to note that, combining the boost $B$ with a translation
on a spectator direction, one obtains the Gott space-time, i.e. the geometry 
around cosmic strings in four dimensions \cite{Gott:1990zr}. 

Due to the compactness of the time coordinate $\eta$, both Misner and
Taub-NUT space-times contain closed timelike curves (CTC) which are usually
considered as a severe pathology. In addition to logical
paradoxes and exciting prospects \cite{lem} raised by time-loops, 
the energy-momentum tensor generated by a scalar field at one-loop is 
typically divergent, indicating a large quantum 
back-reaction. According to the  Chronology Protection Conjecture, this
back-reaction may prevent the formation of CTC 
altogether \cite{Hawking:1991nk}. String theorists
need not be intimidated by such considerations, and boldly go and investigate whether
the magics of string theory alleviate some of these problems.

String theory on a quotient of flat space\footnote{String theory on Taub-NUT
space, which is not flat, has been studied recently using heterotic 
coset models \cite{Johnson:2004zq}.} 
is in principle amenable to standard orbifold
conformal field theory techniques, although the latter are usually formulated
for Euclidean orbifolds. While backgrounds with Lorentzian signature 
can often be dealt with by (often subtle) Wick rotation from Euclidean 
backgrounds, the real complication stems from the fact that the orbifold
group is infinite, and its action non proper \footnote{Defining $X^+ = Z\ ,
\quad X^- = -\bar Z$ in the Rindler
region, one obtains an orbifold of $\mathbb{R}^2$ by a rotation with an irrational
angle. A related model has been studied recently in  \cite{Kutasov:2004aj}.}.
This however  need not be a problem at a classical level: as we shall see, 
free strings propagate in a perfectly well-defined fashion on this 
singular geometry. 

\subsection{Particles in Misner space}

As in standard orbifolds, part of the closed 
string spectrum consists of configurations on the covering space, which 
are invariant under the orbifold action. Such ``untwisted'' states behave
much like point particles of arbitrary mass and spin. Their trajectory,
aside from small-range string oscillations, consist of straight lines
on the covering space: 
\begin{equation}
\label{clasu} X_0^\pm =
x_0^\pm + p^\pm \tau \ ,
\end{equation}
where $m^2 = 2 p^+ p^-$ includes the contribution from momentum in the 
transverse
directions to Misner space as well as string oscillators. 
The momentum along the
compact direction is the ``boost momentum'' $j=x_0^+ p^- + x_0^- p^+$, and
is quantized in units of $1/\beta$ in the quantum theory. 
A massive particle with positive energy ($p^+,p^->0$) thus 
comes in from the infinite past in
the Milne region at $\tau=-\infty$ and exits in the
future Milne region at $\tau=+\infty$, after wandering in the Rindler regions
for a finite proper time. As the particle approaches the light-cone from the
past region, its angular velocity $d\theta/dT \sim 1/T$ along the Milne circle 
increases to infinity by the familiar ``spinning skater'' effect. It is 
therefore expected to emit abundant gravitation radiation, and possibly
lead to large back-reaction. From the point of view of an observer in
one of the Rindler regions, an infinite number
of particles of Rindler energy $j$ are periodically emitted 
from the horizon at $r=0$ 
and travel up to a finite radius $r=|j|/M$ 
before being reabsorbed into the singularity
-- and so on around the time loop.

Quantum mechanically, the center of mass of a (spinless) untwisted string 
is described by a wave function, solution of the Klein-Gordon equation in
the Misner geometry. Diagonalizing the boost momentum $j$, the radial motion
is governed by a Schr\"odinger equation
\begin{equation}
\left\{ 
\begin{matrix} -\partial_x^2 - m^2 e^{2y} - j^2 = 0 \\
-\partial_y^2 + m^2 e^{2y} - j^2 = 0  \end{matrix}
\right.
\quad \mbox{where} \quad
\left\{ 
\begin{matrix} T&=&\pm\sqrt{2x^+ x^-}=e^{x}\\
r&=&\pm\sqrt{-2x^+ x^-}=e^{y}\\
\end{matrix}
\right.
\end{equation}
The particle therefore bounces against an exponentially rising,
Liouville-type wall in the Rindler regions, while it is accelerated in a 
Liouville-type well in the Misner regions. Notice that, in both cases, the 
origin lies at infinite distance in the canonically normalized 
coordinate $x$ ($y$, resp.). Nevertheless, $in$ and $out$
type of wave functions can be defined in each region, and extended
to globally defined wave functions
by analytic continuation across the horizons at $x^+ x^-=0$.

Equivalently,
the wave function for an untwisted string 
in Misner space may be obtained by superposing 
a Minkowski plane wave with its images under the
iterated boosts $B^n,\ n\in\mathbb{Z}$. Performing a Poisson resummation over 
the integer $n$, one obtains wave functions with a well defined value
of the boost momentum $j$, as a continuous superposition of plane waves
\begin{equation} \label{propoco}
f_{j,m^2,s}(x^+,x^-)=\int_{-\infty}^{\infty}dv\ 
\exp\left( i p^+ X^- e^{-2\pi\beta v}
+ i p^- X^+ e^{2\pi\beta v} + i v j + v s \right)
\end{equation}
where $s$ denotes the $SO(1,1)$ 
spin in $\mathbb{R}^{1,1}$ \cite{Nekrasov:2002kf,Berkooz:2004re}. 
This expression defines global wave functions in all regions,
provided the $v$-integration contour is deformed to 
$(-\infty-i\epsilon,+\infty+i\epsilon)$. In particular, there is no overall particle
production between the (adiabatic) $in$ vacuum at $T=-\infty$ and the $out$
vacuum at $T=+\infty$, however there is particle production between the
(adiabatic) $in$ vacuum at $T=-\infty$ and the (conformal) $out$ vacuum
at $T=0^-$. This is expected, due to the ``spinning skater'' 
infinite acceleration near the singularity, as mentioned above.

\subsection{Winding strings in Misner space}

In addition to the particle-like untwisted states, the orbifold spectrum
contains string configurations which close on the covering space, up to
the action of an iterated boost $B^w$:
\begin{equation}
\label{perw}
X^{\pm}(\sigma+2\pi,\tau)=e^{\pm 2\pi w \beta} X^{\pm}(\sigma,\tau)
\end{equation}
In the Milne
regions,  they correspond to strings winding $w$ times
around the compact space-like 
dimension $S^1_\theta$, which become massless at the cosmological
singularity. They are therefore expected to play a prominent r\^ole
in its resolution, if at all. In the Rindler regions, they instead 
correspond to strings winding around the compact 
{\it time-like} dimension $S^1_\eta$. Given that a time-loop exist,
there is nothing {\it a priori} wrong about a string winding around time:
it is just a superposition of $w$ static 
(or, more generally, periodic in time) strings,
stretched (in the case of a cylinder topology)
over an infinite radial distance. 

In order to understand the semi-classical aspects 
of twisted strings \cite{Berkooz:2004re},
let us again truncate to the modes with lowest worldsheet energy,
satisfying \eqref{perw}:
\begin{equation}
\label{x0}
 X^{\pm}_0(\sigma,\tau) =
\frac1{\nu} e^{\mp \nu \sigma} \left[ \pm
\alpha^\pm_0 e^{\pm \nu \tau} 
\mp \tilde \alpha_0^\pm e^{\mp \nu \tau}\right]\ .
\end{equation}
where $\nu=-w\beta$.
As usual, the Virasoro (physical state) conditions determine the 
mass and momentum of the state in terms of the oscillators,
\begin{equation}
\label{tvir}
M^2 = 2 \alpha_0^+ \alpha_0^-   \ ,\quad
\tilde M^2 =  2 \tilde\alpha_0^+ \tilde\alpha_0^- 
\end{equation}
where $(M^2=m^2+j\nu,\tilde M^2=m^2-j\nu)$ are the contributions of the
left-moving (resp. right-moving) oscillators.
Restricting to $j=0$ for simplicity, one may thus choose 
$\alpha^\pm$ and $\tilde\alpha^\pm$ to be all equal in modulus to 
$m/\sqrt2$, up to choices of sign leading to two qualitatively
different kinds of twisted strings:
\begin{itemize}
\item For $\alpha^+\tilde\alpha^->0$, one obtains {\it  short string}
configurations
\begin{equation}
X^\pm_0(\sigma,\tau)= \frac{m}{\nu\sqrt2}\ \sinh(\nu\tau)\ 
e^{\pm \nu \sigma}
\end{equation}
winding around the Milne space-like circle, and propagate from infinite past 
to infinite future (for $\alpha^+>0$). 
When $j\neq 0$, they also extend in the Rindler regions 
to a finite distance $r_-^2=(M-\tilde M)^2/(4\nu^2)$, after experiencing
a signature flip on the worldsheet.
\item For $\alpha^+\tilde\alpha^-<0$, one obtains {\it  long string}
configurations, 
\begin{equation}
\label{x0t}
X^\pm_0(\sigma,\tau)= \frac{m}{\nu\sqrt2}\ \cosh(\nu\tau)\ 
e^{\pm \nu \sigma}
\end{equation}
propagating in the Rindler regions only, and winding
around the time-like circle. 
They  correspond to static configurations which extend from spatial
infinity in L or R to a finite distance $r_+^2=(M+\tilde M)^2/(4\nu^2)$,
and folding back to infinity again.
\end{itemize}
Notice how, in contrast to Euclidean orbifolds, twisted strings
are in no sense localized near the singularity !

Quantum mechanically, the (quasi) zero-modes $\alpha_0^\pm, \tilde\alpha_0^\pm$
become hermitian operators with commutation rules  \cite{Nekrasov:2002kf,Pioline:2003bs}
\begin{equation}
\label{com0}
[\alpha_0^+, \alpha_0^-]= -i\nu\ ,\quad
[\tilde \alpha_0^+, \tilde \alpha_0^-]= i\nu
\end{equation}
Representing $\alpha_0^+$  as a creation operator in a Fock space
whose vacuum is annihilated by $\alpha_0^-$, introduces an imaginary
ordering constant $i\nu/2$ in \eqref{tvir} after normal ordering, 
which cannot be cancelled by any of the higher modes in the 
spectrum\footnote{Higher excited modes have energy $n\pm i\nu$,
and can be quantized in the usual Fock space scheme,
despite the Lorentzian signature of the light-cone directions
 \cite{Pioline:2003bs}.}.
Thus, in this scheme, there are no physical states in the twisted
sector \cite{Nekrasov:2002kf}. However, this quantization does not
maintain the hermiticity of the zero-mode operators. The analogy
of \eqref{com0} to the problem of a charged particle in an electric field
will take us to the appropriate quantization scheme in the next section.

\subsection{Winding strings vs. charged particles}
\label{ws}

\begin{figure}
\begin{center}
\epsfig{file=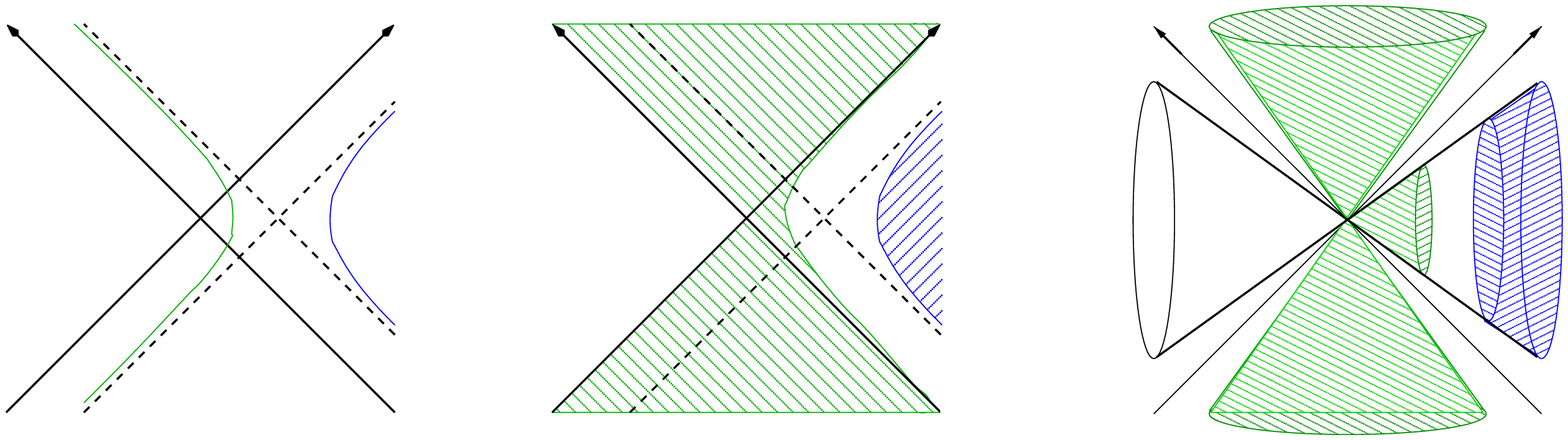,height=3cm}
\caption{\label{cutting} Closed string worldsheets in Misner space are obtained by smearing
the trajectory of a charged particle in Minkowski space with a constant 
electric field. Short (resp. long) strings correspond to charged particles
which do (resp. do not) cross the horizon.}
\end{center}
\end{figure}

Returning to \eqref{x0t}, one notices that the complete worldsheet of a 
twisted closed string can be obtained by smearing the trajectory of the
left-movers (i.e. a point with $\tau+\sigma=cste$) under the action
of continuous boosts (See Figure \ref{cutting}). In particular, setting $a_0^\pm = \tilde\alpha_0^\pm$
and $x_0^\pm = \mp \tilde\alpha_0^\pm/ \nu$, the trajectory of the left-movers
becomes 
\begin{equation}
X^\pm(\tau) = x_0^\pm \pm \frac{a_0^\pm}{\nu}
e^{\pm \nu \tau}\ .
\end{equation}
which is nothing but the worldline of a particle of charge $w$ in a
constant electric field $E=\beta$ ! Indeed, it is easily verified that 
the short (long, resp.) string worldsheet can be obtained by smearing the 
worldline of a charged particle which crosses (does not, resp.) 
the horizon at $x^+ x^-=0$. 

Quantum mechanically, it is easy to see that this analogy continues to 
hold \cite{Pioline:2003bs,Berkooz:2004re} : 
the usual commutation relations for a particle in an electric field
\begin{equation}
[a_0^+, a_0^-]= -i\nu\ ,\quad
[x_0^+, x_0^-]= -\frac{i}{\nu}\ ,\quad
\end{equation}
reproduce the closed string relations \eqref{com0} under the identification
above. The mass of the charged particle $M^2=\alpha^+\alpha^-+\alpha^-\alpha^+$
reproduces the left-moving Virasoro generator $m^2 + \nu j$ as well.
It is therefore clear that the closed string zero-modes, just as their
charge particle counterpart, can be represented as covariant derivatives
acting on complex wave functions $\phi(x^+,x^-)$:
\begin{equation} 
\label{reals} \alpha_0^\pm= i \nabla_{\mp} = i \partial_\mp
\pm \frac{\nu}{2} x^\pm\ ,\quad \tilde\alpha_0^\pm= i
\tilde \nabla_{\mp} = i \partial_\mp \mp \frac{\nu}{2}
x^\pm
\end{equation} 
in such a way that the physical state
conditions are simply the Klein-Gordon operators
for a particle with charge $\pm \nu$ in a constant
electric field,
\begin{equation}
M^2= \nabla_+ \nabla_- +
\nabla_- \nabla_+\ ,\quad \tilde M^2= \tilde\nabla_+
\tilde\nabla_- + \tilde\nabla_- \tilde\nabla_+
\end{equation}
Coordinates $x^\pm$ are the (Heisenberg picture)
operators corresponding to
the location of the closed string at $\sigma=0$.
The radial coordinate $\sqrt{\pm 2 x^+x^-}$ associated
to the coordinate representation \eqref{reals}, should be
thought of as the radial position of the closed
string in the Milne or Rindler regions.

From this point of view, it is also clear while the quantization
scheme based on a Fock space has failed: the Klein-Gordon equation
of a charged particle in an electric field is equivalent, for
fixed energy $p_t$, to a Schr\"odinger equation with 
an {\it inverted} harmonic potential,
\begin{equation}
-\partial_x^2 + m^2 - (p_t - E x)^2 \equiv 0
\end{equation}
In contrast to the magnetic case which leads to 
a positive harmonic potential with discrete Landau levels, 
the spectrum consists of a continuum of delta-normalization 
scattering states which bounce off (and tunnel through) the potential barrier.
These scattering states are the quantum wave functions
corresponding to electrons and positrons being reflected by the 
electric field, and their mixing under tunneling is a reflection of 
Schwinger production of charged pairs from the vacuum. 

In order to apply this picture to twisted closed strings however, we 
need to project on boost invariant states, and therefore understand
the charged particle problem from the point of view of an accelerated
observer in Minkowski space, i.e. a static observer in Rindler space.

\subsection{Charged particles in Misner space}

\begin{figure}
\begin{center}
\epsfig{file=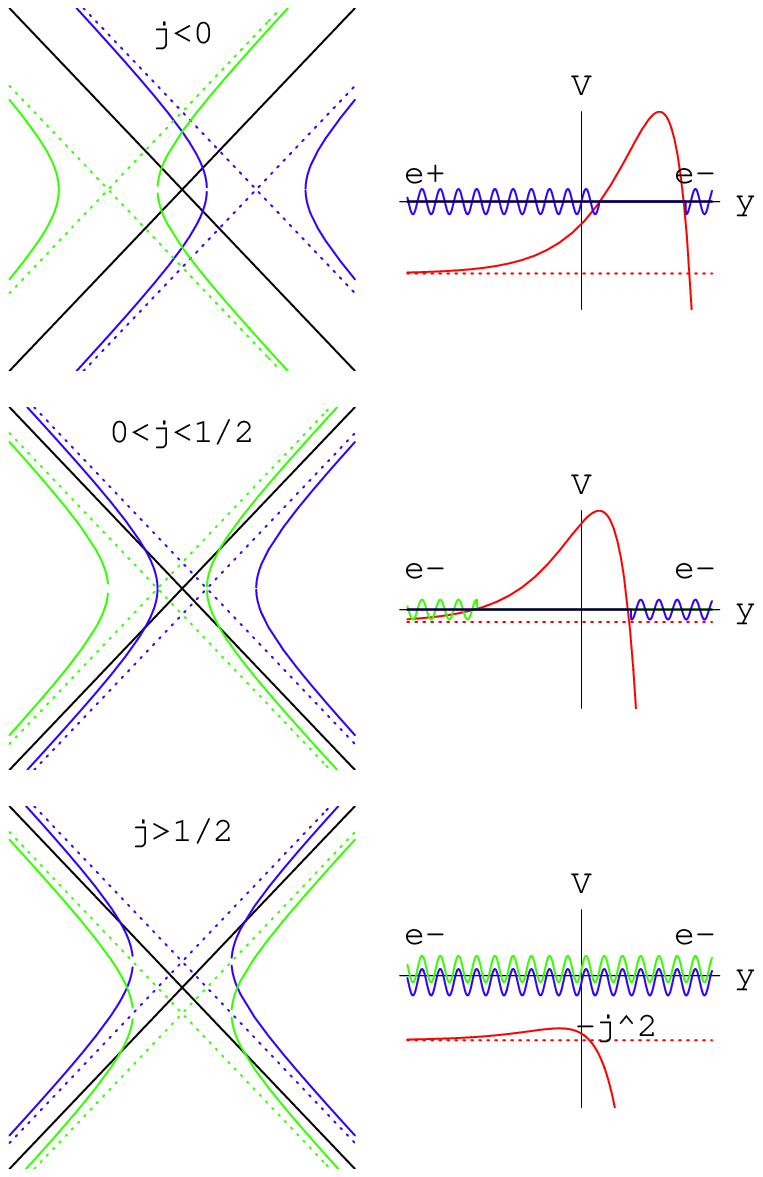,height=14cm}
\end{center}
\caption{\label{trajclas}}
{\small Left: Classical trajectories of a charged particle in Rindler/Milne space.
$j$ labels the Rindler energy or Milne momentum, and is measured in units
of $M^2/\nu$. Right: Potential governing the radial motion in the right
Rindler region, as a function of the canonical coordinate $y=e^r$.}
% \caption{Left: Classical trajectories of a charged particle in Rindler/Milne space.
% $j$ labels the Rindler energy or Milne momentum, and is measured in units
% of $M^2/\nu$. Right: Potential governing the radial motion in the right
% Rindler region, as a function of the canonical coordinate $y=e^r$. }
\end{figure}

Charged particles in Rindler space have been discussed in 
\cite{Gabriel:1999yz}.
%, Cooper:1992hw, Narozhny:2003ux}.
Classical trajectories are, of course, the ordinary hyperbolae 
from Minkowski space, translated into the Rindler coordinates $(y=e^r,\eta)$.
For a fixed value $j$ of the energy conjugate to the Rindler time $\eta$,
the radial motion is governed by the potential
\begin{equation} \label{schror} V(y) = M^2 r^2 -
\left( j + \frac12 \nu r^2 \right)^2 = \frac{M^2
\tilde M^2}{\nu ^2} - \left( \frac{M^2+\tilde M^2}{2\nu} -
\frac{\nu}{2} r^2 \right)^2
\end{equation}
where, in the last equality, we have translated the charged particle
data into closed string data. In contrast with the neutral case ($\nu=0$),
the potential is now unbounded from below at $r=\infty$. For $j<M^2/(2\nu)$
(which is automatically obeyed in the closed string case, where 
$\tilde M^2>0$ for non-tachyonic states), the $r=0$ and $r=\infty$ 
asymptotic regions are separated by a potential barrier (See Figure \ref{trajclas}). 
Particles on
the right ($r\to\infty$) of the barrier correspond to electrons coming
from and returning to Rindler infinity, while, for $j>0$ (resp. $j<0$),
particles on the left ($r\to 0$) correspond to positrons (resp. electrons)
emitted from and reabsorbed by the Rindler horizon. Quantum tunneling
therefore describes both Schwinger pair production in the electric
field (when $j>0$), and Hawking emission of charged particles 
from the horizon (when $j<0$). 

Similarly, the trajectories of charged particles in Milne space correspond 
to other branches of the same hyperbolae, and their motion along the 
cosmological time $T$, for a fixed value of the momentum $j$ conjugate
to the compact spatial direction $\theta$, is governed by the potential
\begin{equation}
 \label{schrom} V(T) = - M^2 T^2 -
\left( j + \frac12 \nu T^2 \right)^2 = \frac{M^2
\tilde M^2}{\nu ^2} - \left( \frac{M^2+\tilde M^2}{2\nu} +
\frac{\nu}{2} T^2 \right)^2
\end{equation}
The potential is maximal and negative at $T=0$, 
although this is at infinite distance
in the canonically normalized coordinate $x$. The classical 
motion therefore covers the complete time axis $T\in\mathbb{R}$.

Quantum mechanically, the Klein-Gordon equation in the Rindler region is 
equivalent to a Schr\"odinger equation in the potential \eqref{schror}
or \eqref{schrom} at zero-energy, and can be solved in terms of 
Whittaker functions \cite{Gabriel:1999yz}. Bases of {\it in} and {\it out} modes
can be defined in each quadrant and analytically continued accross the
horizons, e.g. in the right Rindler region
\begin{equation}
\label{vinr}
{\cal V}_{in,R}^j = e^{-i j \eta} r^{-1}
M_{-i(\frac{j}{2}-\frac{M^2}{2\nu}), -\frac{ij}{2} } ( i \nu r^2 /2 )
\end{equation}
corresponds to incoming modes from Rindler infinity $r=\infty$, while
\begin{equation}
\label{uinr}
{\cal U}_{in,R}^j = e^{-i j \eta} r^{-1}
W_{i(\frac{j}{2}-\frac{M^2}{2\nu}), \frac{ij}{2} } ( -i \nu r^2 /2 )
\end{equation}
corresponds to incoming modes from the Rindler horizon $r=0$.
As usual in time-dependent backgroungs, the $in$ and $out$ vacua are
related by a non-trivial Bogolubov transformation, which implies
that production of correlated pairs has taken place. The
Bogolubov coefficients have been computed in \cite{Gabriel:1999yz,
Pioline:2003bs}, and yield the pair creation rates in the
Rindler and Milne regions, respectively:
\begin{equation}
\label{ref}
Q_{R} =
 e^{-\pi M^2/2\nu}
\frac{|\sinh \pi j|}{\cosh\left[ \pi \tilde M^2/ 2\nu\right]}\ ,\quad
Q_{M} =
 e^{-\pi M^2/2\nu}
\frac{\cosh\left[ \pi \tilde M^2/ 2\nu\right]}{|\sinh \pi j|}\ ,
\end{equation}
In the classical limit $M^2,\tilde M^2 \gg \nu$, these indeed agree
with the tunnelling (or scattering over the barrier, in the
Milne regions) rate computed from \eqref{schrom}.

\subsection{Schwinger pair production of winding strings}
Having understood the quantum mechanics of charged particles in Minkowski
space from the point of view of an accelerating observer, we 
now return to the dynamics of twisted strings in Misner space.
The wave function of the quasi-zero-modes $\alpha_0^\pm,\alpha_0^\pm$
is governed by the same Klein-Gordon equation as in the charged particle
case, although only the dependence on the radial coordinate $r$ is of 
interest. Its interpretation is however rather different: e.g, 
a particle on the right of the potential in the right Rindler region
corresponds to an infinitely long string stretching from infinity in 
the right whisker to
a finite radius $r_+$ and folded back onto itself, while 
a particle on the left of the potential is a short string stretching
from the singularity to a finite radius $r_-$. Quantum tunneling relates
the two type of states by evolution in imaginary radius $r$, and can be
viewed semi-classically as an Euclidean strip stretched between $r_+$ 
and $r_-$. The wave functions in the Milne region are less exotic,
corresponding to incoming or outgoing short strings at infinite past, future
or near the singularity. 

In order to compute pair production, one should in principle define 
second quantized vacua, i.e. choose a basis of positive and negative
energy states. While it is clear how to do so for short strings 
in the Milne regions, second quantizing long strings is less evident,
as they carry an infinite Rindler energy\footnote{The latter can be 
computed by quantizing the long string worldsheet using $\sigma$
as the time variable \cite{Berkooz:2004re}.}, and depend on the boundary conditions 
at $r=\infty$. However, they are likely to give the most natural formulation, as 
any global wave function in Misner space
can be written as a state in the tensor product 
of the left and right Rindler regions: the entire cosmological dynamics 
may thus be described as a state in a time-independent geometry, albeit 
with time loops ! 

Fortunately, even without a proper understanding
of these issues, one may still use the formulae \eqref{ref} to relate
incoming and outcoming components of the closed string wave functions,
and compute pair production for given boundary conditions at Rindler
infinity. In particular, it should be noted that the production rate
in the Milne regions $Q_M$ is infinite for vanishing boost momentum $j=0$,
as a consequence of the singular geometry.

Moreover, although our analysis has borrowed a lot of intuition from
the analogy to the charged particle problem, we are now in a position
to describe pair production of winding strings in any geometry of the
form 
\begin{equation}
ds^2 = -dT^2 + a^2(T) d\theta^2 \quad \mbox{or} \quad
ds^2 = dr^2 - b^2(r) d\eta^2
\end{equation}
(despite the fact that these geometries are not exact solutions of the
{\it tree level} string equations of motion, they may be a useful mean
field description of the back-reacted geometry).
Neglecting the contributions of excited modes (which no longer decouple
since the metric is not flat), the wave equation for the
center of motion of strings winding around the compact direction 
$\theta$ or $\eta$, is obtained by adding to the two-dimensional Laplace 
operator describing the free motion of a neutral particle, the 
contribution of the tensive energy
carried by the winding string:
\begin{equation}
\left\{ \begin{matrix}
\frac{1}{a(T)}\partial_T\ a(T)\ \partial_T
+ \frac{j^2}{a^2(T)} + \frac14 w^2 a^2(T) - m^2 =0 \\
\frac{1}{b(r)}\partial_r\ b(r)\ \partial_r
+ \frac{j^2}{b^2(r)} + \frac14 w^2 b^2(r) - m^2 =0 
\end{matrix}
\right.
\end{equation}
Choosing $a(T)=\beta T$ or $a(r)=\beta r$ and multiplying out by $(a^2(T),
b^2(r))$, these equations indeed
reduce to \eqref{schrom} and \eqref{schror}\footnote{Notice that, in 
disagreement to a claim in the literature \cite{Turok:2004gb}, the wave 
equation for $j=0$ is {\it not} regular at the origin.}. In particular,
for a smooth geometry, the production rate of pairs of winding
strings is finite.

\section{One-loop vacuum amplitude}
In the previous section, we have obtain the production
rate of winding strings in Misner space, from the Bogolubov
coefficients of the tree-level wave functions. In principle,
the same information could be extracted from the imaginary part of
the one-loop amplitude. In this section, we start by reviewing
the vacuum amplitude and stress-energy in field theory, and
go on to study the one-loop amplitude in string theory, both
in the twisted and untwisted sectors.

\subsection{Field theory}
The one-loop energy-momentum tensor generated by the
quantum fluctuations of a free field $\phi$ with
(two-dimensional) mass $M^2$ and spin $s$ can be derived from the 
Wightman functions at coinciding points (and derivatives thereof). 
These depend on the choice of vacuum: in the simplest ``Minkowski''
vacum inherited from the covering space, 
any Green function is given by a sum over images of the
corresponding one on the covering space. Using a
(Lorentzian) Schwinger time representation and integrating over
momenta, we obtain
\begin{eqnarray}
\label{propco} G(x^\mu;{x'}^\mu)=&&
\sum_{l=-\infty}^{\infty}\int_0^\infty d\rho
\ (i\rho)^{-D/2} \exp\left[ i \rho M^2
-2\pi s l \right] \\
&& \exp\left[ - \frac{i}{4\rho}(x^+ - e^{2\pi\beta l}
x^{+'})(x^- - e^{-2\pi\beta l}
x^{-'}) \right] \nonumber\end{eqnarray} 
where $s$ is the total spin carried by the field
bilinear. Taking two derivatives and setting $x=x'$,
one finds a divergent stress--energy tensor \cite{Hiscock:vq}
\begin{equation}
T_{\mu\nu} dx^{\mu}dx^{\nu} = \frac{1}{12\pi^2} \frac{K}{T^4}
\left( -dT^2 -3 T^2 d\eta^2 \right)
\end{equation}
where the constant $K$ is given by 
\begin{equation} K= \sum_{n=1}^{\infty}
\cosh[2\pi n s \beta] \frac{2+\cosh 2\pi n
\beta}{(\cosh 2\pi n \beta-1)^2} 
\end{equation}
This divergence is expected due to the large blue-shift of quantum 
fluctuations near the singularity.
Notice that for spin $|s|>1$, the constant $K$
itself becomes infinite, a reflection of the
non-normalizability of the wave
functions for fields with spin.

In string theory, the local expectation value
$\langle 0  | T_{ab}(x) | 0 \rangle_{ren}$ is not
an on-shell quantity, hence not directly
observable. In contrast, the integrated free
energy, given by a torus amplitude, is a valid
observable\footnote{Of course, the spatial dependence
of the one-loop energy may be probed by scattering
e.g. gravitons at one-loop.}. In field theory,
the free energy may be obtained by integrating
the propagator at coinciding points \eqref{propco}
once with respect to $M^2$, as well as over all
positions, leading to
\begin{equation} \label{ftg}
\mathcal{F} = \sum_{l=-\infty}^{\infty} \int dx^+
dx^- \int_0^\infty
\frac{d\rho}{(i\rho)^{1+\frac{D}{2}}} \exp\left( -8
i \sinh^2 (\pi \beta l) x^+ x^- +
i M^2 \rho  \right) 
\end{equation} In contrast to the flat
space case, the integral over the zero-modes
$x^\pm,x$ does not reduce to a volume factor, but
gives a Gaussian integral, centered on the light
cone $x^+ x^-=0$. Dropping as usual the divergent
$l=0$ flat-space contribution and rotating to imaginary
Schwinger time, one obtains a finite
result \begin{equation} \label{ft} \mathcal{F} =
\sum_{l=-\infty, l\neq 0}^{+\infty} \int_0^{\infty}
\frac{d\rho}{\rho^{1+\frac{D}{2}} }\frac{e^{- M^2
\rho -2\pi \beta s
l} } {\sinh^2\left( \pi \beta l\right)}
\end{equation} 
Consistently with the existence of globally defined 
positive energy modes for (untwisted) particles in Misner
space, ${\cal F}$ does not have any
imaginary part, implying  the absence of net
particle production between  past and future
infinity. 

\subsection{String amplitude in the untwisted sector}
We may now compare the field theory result
\eqref{ft} to the one-loop vacuum amplitude in
string theory  with Euclidean world-sheet and
Min\-kowsk\-ian target space, as computed in
\cite{Nekrasov:2002kf,Cornalba:2002fi}: \begin{equation}
\label{milne1l} A_{bos}=\int_{{\cal F}}
\sum_{l,w=-\infty}^{\infty} \frac{d\rho
d\bar\rho}{(2\pi^2 \rho_2)^{13}} \frac{e^{-2\pi
\beta^2 w^2 \rho_2 -
\frac{R^2}{4\pi\rho_2}|l+w\tau|^2}} {\left|
\eta^{21}(\rho) ~\theta_1(i \beta(l+w\rho); \rho)
\right| ^2 } \end{equation} where $\theta_1$ is the Jacobi
theta function, \begin{equation} \label{jac} \theta_1(v;\rho)= 2
q^{1/8} \sin \pi v \prod_{n=1}^\infty (1-e^{2\pi i
v} q^n) ( 1-q^n) (1-e^{-2\pi i v} q^n) \ ,\quad
q=e^{2\pi i\rho} \end{equation} In this section, we restrict
to the untwisted sector $w=0$. Expanding in powers
of $q$, it is apparent that the string theory
vacuum amplitude can be viewed as the field theory
result \eqref{ft} summed over the spectrum of
(single particle) excited states, satisfying the
matching condition enforced by the integration over
$\rho_1$. As usual, field-theoretical UV
divergences at $\rho\to 0$ are cut-off by
restricting the integral to the fundamental domain
$F$ of the upper half plane.

In contrast to the field theory result, where the
integrated free energy is finite for each particle
separately, the free energy here has  poles in the
domain of integration, at
\begin{equation} \rho= \frac{m}{n} + i
\frac{\beta l}{2\pi n} \label{dp}\end{equation}
Those poles arise only
after summing over infinitely many  string theory
states. Indeed, each pole originates from the
$(1-e^{\pm 2\pi i v} q^n)$ factor in \eqref{jac},
hence re-sums the contributions of a complete Regge
trajectory of fields with mass $M^2= kn$ and spin
$s=k$ ($k\in\mathbb{Z}$). In other words, the usual
exponential suppression of the partition function
by the increasing masses along the Regge trajectory
is overcome by the spin dependence of that
partition function. Regge trajectories are a
universal feature of perturbative string theory,
and these divergences are expected generically in the presence
of space-like singularities. Since the pole \eqref{dp} occurs
both on the left- and right-moving part, the integral is not expected
to give any imaginary contribution, in contrast to the 
charged open string case considered in \cite{Bachas:bh}.

\subsection{String amplitude in the twisted sectors}
We now turn to the interpretation of the string one-loop amplitude
 in the twisted sectors ($w\neq 0$), following the analysis in  
\cite{Berkooz:2004re}. As in the rest of this
lecture, it is useful to truncate the twisted string to its 
quasi-zero-modes, lumping together the excited mode contributions into
a left and right-moving mass squared $M^2$ and $\tilde
M^2$. Equivalently, we truncate the path integral to 
the ``mini-superspace'' of lowest energy 
configurations on the torus of modulus $\rho=\rho_1+i \rho_2$, 
satisfying the twisted boundary conditions,
\begin{equation} \label{qzm}
X^\pm = \pm \frac{1}{2\nu} \alpha^\pm e^{\mp (\nu
\sigma - i A \tau)} \mp \frac1{2\nu} \tilde\alpha^\pm
e^{\mp (\nu\sigma + i \tilde A\tau)}
\end{equation}
where
\begin{equation}
A= \frac{k}{\rho_2} - i
\beta \frac{l + \rho_1 w}{\rho_2}\ ,\quad 
\tilde
A= \frac{\tilde k}{\rho_2} + i \beta \frac{l +
\rho_1 w}{\rho_2}
\end{equation}
where $k,\tilde k$ are a pair of
integers labelling
the periodic trajectory, for fixed twist numbers $(l,w)$.
Notice that \eqref{qzm} is not a solution of the
equations of motion, unless $\rho$ coincides with
one of the poles. In order to satisfy the
reality condition on $X^\pm$, one should restrict
to configurations with $k=\tilde k$,
$\alpha^\pm = - (\tilde\alpha^\pm)^*$.
Nevertheless, for the sake of generality we shall not
impose these conditions at this stage, but only exclude the case of 
a degenerate worldsheet $k=-\tilde k$.

We can now evaluate the Polyakov action for such a
classical configuration, after rotating $\tau\rightarrow i\tau$:
\begin{equation}\label{ea}
\begin{split}
S =& -\frac{\pi }{2\nu^2 \rho_2}
\left( \nu^2 \rho_2^2 - \left[k - i (\beta l + \nu
\rho_1)\right]^2 \right) R^2 \\
&  -\frac{\pi }{2\nu^2 \rho_2}
\left( \nu^2 \rho_2^2 - \left[\tilde k + i (\beta l + \nu \rho_1)\right]^2
\right)\tilde R^2 \\
& \hspace*{2cm} - 2 \pi i j \nu \rho_1 + 2 \pi  \mu^2 \rho_2
\end{split}
\end{equation}
where the last line, equal to $-i\pi \rho M^2 +i \pi \tilde \rho \tilde M^2$,
summarizes the contributions of excited modes, and
$\alpha^\pm = \pm R e^{\pm \eta}/\sqrt{2}$,
$\tilde\alpha^\pm = \pm \tilde R e^{\pm \tilde \eta}/\sqrt{2}$.
%$R^2=2\alpha^+\alpha^-, \tilde R^2=2\tilde\alpha^+\tilde\alpha^-$.

The path integral is thus truncated to an integral over
the quasi-zero-modes $\alpha^\pm,\tilde\alpha^\pm$. Since
the action \eqref{ea} depends only on the boost-invariant
products $R^2$ and $\tilde R^2$, a first divergence arises from
the integration over $\eta-\tilde\eta$, giving an infinite factor,
independent of the moduli, while the integral over $\eta+\tilde\eta$
is regulated to the finite value $\beta$ after dividing out 
by the (infinite) order of the orbifold group.

In addition there are divergences coming from the
integration over $R$ and $\tilde R$ whenever
\begin{equation}
\label{r12}
\rho_1=-\frac{\beta l}{\nu} - i \frac{\nu}{2} (k - \tilde k)\ ,\qquad
\rho_2 = \frac{|k+\tilde k|}{2\nu}
\end{equation}
which, for $k=\tilde k$, are precisely the double poles \eqref{dp}.
These poles are interpreted as coming from infrared divergences
due to existence of modes with arbitrary size $(R,\tilde R)$. For $k\neq
\tilde k$, the double poles are now in the
complex $\rho_1$ plane, and may contribute for specific
choices of integration contours, or second-quantized vacua.
In either case, these
divergences may be regulated by enforcing a cut off
$|\rho-\rho_0|>\epsilon$ on the moduli space, or an infrared cut-off on $R$.
It would be interesting to understand the deformation of Misner
space corresponding to this cut off, analogous to the Liouville
wall in $AdS_3$ \cite{Maldacena:2000kv}.

Rather than integrating over $R,\tilde R$ first, which is ill-defined at
$\rho$ satisfying \eqref{r12}, we may choose to integrate
over the modulus $\rho$ first. The integral with respect to
$\rho_1$ is Gaussian, dominated by a saddle point at
\begin{equation}
\rho_1 = - \frac{\beta l}{\nu}+ i \frac{\tilde k \tilde R^2
- k R^2}{\nu (R^2+ \tilde R^2)}
- 2 i \frac{j \nu \rho_2}{R^2+\tilde R^2}
\end{equation}
It is important to note that this saddle point is a local extremum
of the Euclidean action, unstable under perturbations of $\rho_1$.
The resulting Bessel-type action has again
a stable saddle point in $\rho_2$,
at \begin{equation} \rho_2 = \frac{R \tilde R
|k + \tilde k|} {\nu
\sqrt{(R^2+\tilde R^2)(4\mu^2-R^2-\tilde R^2)-4 j^2\nu^2}}
\end{equation}
Integrating over $\rho_2$ in the saddle point
approximation, we finally obtain the action as a
function of the radii $R,\tilde R$: \begin{eqnarray} \label{s2}
S&=&\frac{|k+\tilde k| R\tilde R
  \sqrt{(R^2+\tilde R^2)(R^2+\tilde R^2-4\mu^2)+4j^2\nu^2}}
{\nu(R^2+\tilde R^2)} \nonumber
\\ &&\pm 2\pi  j \frac{\tilde k \tilde R^2-k R^2}{\nu(R^2+\tilde R^2)}
\pm 2\pi i \beta j l
\end{eqnarray}
where the sign of the second term is that of $k+\tilde k$.
$S$ admits an extremum at
the on-shell values
\begin{equation}
\label{sos}
R^2=\mu^2-j\nu, \qquad \tilde R^2=\mu^2+j\nu
\qquad \mbox{with action} \quad
S_{k,\tilde k}= \frac{\pi M \tilde M }{\nu} |k+\tilde k|
\end{equation}
Notice that these values are consistent with the reality condition,
since the boost momentum $j$ is imaginary in Euclidean
proper time. Evaluating $(\rho_1,\rho_2)$ for the values \eqref{sos},
we reproduce \eqref{r12}, which implies that the integral is
indeed dominated by the region around the double pole.
Fluctuations  in $(\rho_1,\rho_2,R,\tilde R)$ directions around
the saddle point have  signature
$(+,+,-,-)$, hence a positive fluctuation determinant,
equal to $M^2 \tilde M^2$ up to a positive numerical constant.
This implies that the one-loop amplitude in the twisted sectors
does not have any imaginary part, in accordance with the naive
expectation based on the double pole singularities. It is also
in agreement with the answer in the untwisted sectors, where
the globally defined $in$ and $out$ vacua where shown to be
identical, despite the occurence of pair production at
intermediate times.

Nevertheless, the instability of the Euclidean action under
fluctuations of $\rho_1$ and $\rho_2$
indicates that spontaneous pair production
takes place, by condensation of the two unstable modes.
Thus, we find that winding string production
takes place in Misner space, at least
for vacua such that the integration contour picks up contributions
from these states. This is consistent with our discussion of the
tree-level twisted wave functions, where tunneling  in the Rindler
regions implies induced pair production of short
and long strings. The periodic trajectories \eqref{qzm}
describe the propagation across the potential barrier in imaginary
proper time, and correspond to an Euclidean world-sheet interpolating
between the Lorentzian world-sheets of the long and short strings.

%We can then isolate on the one hand poles coming from infrared divergences
%due to existence of modes with arbitrary size (in target space)
%and on the other hand poles that may contribute for specific
%choices of integration contours, or second-quantized vacua.
%In either case, these
%divergences may be regulated by enforcing a cut off
%$|\rho-\rho_0|>\epsilon$ on the moduli space, or an infrared cut-off on $R$.

%It is worth noticing also that the integral with respect to $\rho_1$
%is dominated by a saddle point which is a local maximum of the Euclidean
%action, hence unstable under perturbations. When carefully studying fluctuations,
%one finds that there are two unstable modes whereas the fluctuation determinant is real.
%Therefore despite naive expectation based on singularities in the twisted sectors
%and the agreement with the answer in untwisted sector, there exists
%spontaneous pair production arising from condensation of the two unstable modes.

\section{Tree-level scattering amplitudes}

After this brief incursion into one-loop physics, we now return to the
classical realm, and discuss some features of tree-level scattering
amplitudes. We start by reviewing the scattering of untwisted modes,
then turn to 
amplitudes involving
two twisted modes, which can still be analyzed by Hamiltonian methods.
We conclude with a computation of scattering amplitudes for more than
2 twisted modes, which can be obtained by a rather different approach.
Our presentation follows \cite{Berkooz:2002je,Berkooz:2004yy}.

\subsection{Untwisted amplitudes}

Tree-level scattering amplitudes for untwisted states in the Lorentzian
orbifold are easily deduced from tree-level scattering amplitudes on
the covering space, by the ``inheritance principle'': expressing the
wave functions of the incoming or outgoing states in Misner space 
as superpositions of Minkowski plane waves with well-defined boost 
momentum $j$ via Eq.\ \eqref{propoco} (with spin $s=0$), the tree-level scattering amplitude
is obtained by averaging the standard Virasoro-Shapiro amplitude
\begin{equation}
%\begin{multline}
\label{virsha}
{\cal A}_{Mink}=\delta\left(\sum_i p_i\right)
\frac{\Gamma\left(-\frac{{\alpha'}}4 s \right) 
\ \Gamma\left(-\frac{{\alpha'}}4 t \right)
\ \Gamma\left(-\frac{{\alpha'}}4 u \right)}
{\Gamma\left(1+\frac{{\alpha'}}4 s \right)\ 
\Gamma\left(1+\frac{{\alpha'}}4 t \right)\ 
\Gamma\left(1+\frac{{\alpha'}}4 u \right)}
%\end{multline}
\end{equation}
under the actions of 
continuous boosts $p_i^\pm\to p_i^\pm(v)=e^{\pm \beta v_i} p_i^\pm$,
with weight $e^{i j_i v_i}$, on all (but one) external momenta.
Possible divergences come from the boundary of the parameter space
spanned by the $v_i$, where some of the momenta $p_i^\pm(v)$ become
large. In a general (Gross Mende, ($s,t,u\to\infty$ with $s/t,s/u$
fixed) high energy regime, the Virasoro-Shapiro amplitude is exponentially 
suppressed \cite{Gross:1987kz}
and the integral over the $v_i$ converges. However, there are
also boundary configurations with $s,u\to\infty$ and fixed $t$ where
the Virasoro-Shapiro amplitude has Regge behavior $s^{t}$, in agreement
with the fact that the size of the string at high energy grows like 
$\sqrt{\log s}$. In this regime, using the Stirling approximation to
the Gamma functions in \eqref{virsha}, it is easy to see that the
averaged amplitude behaves as
\begin{equation}
A_{Misner} \sim \int^{\infty} dv\ \exp \left[ v \left( i(j_2-j_4) 
-\frac12 \alpha'(p_1^i-p_3^i)^2 + 1 \right) \right]
\end{equation}
hence diverges for small momentum transfer $(p_1^i-p_3^i)^2\leq 2/\alpha'$ in the
directions transverse to Misner space. There are similar collinear
divergences in the other channels as well, both in the bosonic or
superstring case.

The situation is slightly improved in the case of Grant space
(analogous to the ``null brane'' considered in \cite{lms}), i.e.
when the boost identification is combined with a translation of length $R$ on a
direction $x_2$ transverse to the light-cone: in this case, the boost momentum
is no longer quantized (although the sum $R p_2+\beta j$ still is), and
one can construct wave packets which are regular on the horizon, 
by superposition of states of 
different boost momentum \cite{lms,Cornalba:2002fi}. Collinear
divergences remain, albeit in a reduced range of momentum 
transfer \cite{Berkooz:2004yy},
\begin{equation}\label{smearond}
\left(\vec{p}_1+\vec{p}_3\right)^2
\leq \frac{(\sqrt{1+2{\alpha'} E^2}-1)^2}{({\alpha'}   E)^2}\ ,\qquad E=\frac{\beta}{R}
\end{equation}
As $R\to 0$, this reduces to Misner space case as expected. 

As a matter of fact, these divergences may be traced to large 
tree-level graviton exchange near the singularity, or, in the Grant
space case, near the chronological horizon  \cite{Berkooz:2002je}. 
Collinear divergences can in principle be treated in the eikonal 
approximation, i.e. by resumming an infinite series of ladder diagrams. 
While a naive application of the flat space result \cite{Amati:1987uf}
suggests that this resummation may lead to finite scattering 
amplitudes of untwisted states in Misner space \cite{Cornalba:2003kd},
a consistent treatment ensuring that only boost-invariant gravitons
are exchanged has not been proposed yet, and prevents us from drawing 
a definitive conclusion. More generally, it would 
be extremely interesting to develop eikonal
techniques in the presence of space-like singularities, and re-evaluate
the claim in \cite{Horowitz:2002mw} that a single particle
in Misner space will ineluctably cause the space to collapse.

\subsection{Two-twist amplitudes}
As we reviewed in Section \ref{ws}, the zero-mode wave functions in the
twisted sectors form a continuum of delta-normalizable states with arbitrarily
negative worldsheet energy. In contrast to the standard case of 
twist fields of finite order in Euclidean rotation orbifolds,
twisted states in Misner space should thus be described by a continuum of vertex
operators with arbitrarily negative conformal dimension. While the
conformal field theory of such operators remains ill-understood, 
amplitudes with two twisted fields only can be computed by ordinary
operator methods on the cylinder, in the twisted vacua
at $\tau=\pm\infty$ ~\cite{Berkooz:2004yy}.

\subsubsection*{Stringy fuzziness.}

Vertex operators for untwisted 
states are just a boost-invariant superposition of the ordinary
flat space vertex operators. In order to compute their scattering amplitude
against a twisted string, it is convenient to write them as a normal
ordered expression in the twisted Hilbert space. Since the twisted 
oscillators have an energy $n\pm i\nu$ with $n\in \mathbb{Z}$, normal ordering
gives a different contribution than in the untwisted state,
\begin{equation}
\Delta(\nu)\equiv
[X_{\succ 0}^-, X^+_{\prec 0}]- [X_{>0}^-, X^+_{<0}]
= \psi(1+i\nu)+\psi(1-i\nu)-2\psi(1)
\end{equation}
where $\psi(x)=\sum_{n=1}^{\infty} (x+n)^{-1}=d\log\Gamma(x)/dx$.
In the above equation, $X^{\pm}_{>,<}$ (resp. $X^{\pm}_{\succ,\prec}$) denote 
the positive and negative frequency parts (excluding the (quasi) zero-mode
contributions) of the embedding 
coordinates $X^\pm(\tau,\sigma)$, as defined by the untwisted
(resp. twisted) mode expansion. As a result, the vertex operator
for an untwisted tachyon becomes
\begin{equation}
\label{ff}
:e^{i(k^+ X^-+k^- X^+)}:^{(\text{un.})} 
= 
\exp\left[ - k^+ k^- \Delta(\nu) \right] \ 
:e^{i(k^+ X^-+k^- X^+)}:^{(\nu-\text{tw.})} 
\end{equation}
Such a factor is in fact present for all untwisted states, although the
normal ordering prescription is slightly more cumbersome for excited 
states. Since this normal ordering constant
depends on the winding number $w=-\nu/\beta$, it cannot be reabsorbed
by a field redefinition of the untwisted state, nor of the twisted 
string. Instead, it can be interpreted as the  
form factor acquired by untwisted states
in the background of a twisted string, due to the zero-point 
quantum fluctuations of the winding string. 
The latter polarizes
untwisted string states into a cloud of r.m.s. size $\sqrt{\Delta(\nu)}$.
which, while proportional to $\nu$ at small $\nu$, grows
logarithmically with the winding number, 
\begin{equation}
\Delta(\nu)=2\zeta(3) \nu^2 + O(\nu^4) = 
2\log \nu -\frac{23}{20} + O(\nu^{-2})
\end{equation}
Notice that this logarithmic growth winding can be viewed as 
the T-dual of the Regge growth with energy. It is also interesting
to observe the analogy of the form factor in \eqref{ff} with 
similar factors appearing in non-commutative gauge theories with
matter in the fundamental representation -- in line with the
general relation between twisted strings and charged particles
outlined in Section 1.4.

\subsubsection*{Zero-mode overlaps.}
In general, the S-matrix element factorizes into a product of an
excited mode contribution, which can be evaluated,
just as in flat space, by normal ordering and commutation, 
and a (quasi)-zero-mode contribution. In the real space
representation \eqref{reals} for the quasi-zero-modes, the latter
reduces to an overlap of twisted and untwisted wave functions,
e.g. in the three  tachyon case,
\begin{equation}
\label{zomo}
\int dx^+ dx^- \ f_1^*(x^+,x^-)\ 
e^{i(p_2^- x^++p_2^+ x^-)}\ f_3(x^+,x^-)
\end{equation}
where $f_1$ and $f_3$ denote eigenmodes of the charged Klein-Gordon
equation, and $f_2$ is an eigenmode of the neutral
Klein-Gordon equation, each of which with fixed angular momentum 
$j_i$. Considering higher excited modes such as the graviton would introduce
extra factors of covariant derivatives $\alpha^\pm$ in \eqref{zomo}.

In order to evaluate these overlaps, it is convenient to use a 
different representation and diagonalize half of the covariant 
derivative operators, e.g.
\begin{equation} \alpha^-=i\nu\partial_{\alpha^+},\ \ \
\tilde\alpha^+=i\nu\partial_{\tilde\alpha^-} \label{oscrin}
\end{equation}
acting on functions of the variables $\alpha^+,\tilde\alpha^-$ taking values
in the quadrant $\mathbb{R}^{\epsilon}\times \mathbb{R}^{\tilde\epsilon}$.
On-shell wave  functions are now powers of their arguments,
\begin{equation} \label{eigoin} f( \alpha^+,  \tilde\alpha^- ) = N_{in}
\left( \epsilon ~ \alpha^+ \right)^{\frac{M^2}{2i\nu} -
\frac12} \left( \tilde\epsilon~ \tilde\alpha^-
\right)^{\frac{\tilde M^2}{2i\nu} - \frac12} \end{equation}
The notation $N_{in}$ for the normalization factor anticipates
the fact that this representation
is appropriate to describe an {\it in} state. The choice of the
signs $\epsilon$ and $\tilde\epsilon$ of $\alpha^-$ and $\tilde\alpha^+$
distinguishes between short strings (
$\epsilon\tilde\epsilon=1$) and long strings ($\epsilon\tilde\epsilon=-1$).
Of course, the oscillator representation \eqref{oscrin} can be related
to the real-space representation via the intertwiner
\begin{equation} f(x^+,x^-) =
\int d\tilde\alpha^+ d\alpha^- \Phi^{in}_{ \nu, \tilde\alpha^+,
\alpha^-}(x^+,x^-) f( \alpha^+, \tilde\alpha^-) \end{equation}
where the kernel is given  by
\begin{equation} \label{phiin} \Phi^{in}_{ \nu,
\alpha^+,
\tilde\alpha^-}(x^+,x^-) = \exp\left( \frac{i\nu x^+
x^-}{2} - i \alpha^+ x^- -i \tilde\alpha^- x^+ +
\frac{i}{\nu} \alpha^+ \tilde\alpha^- \right) \end{equation}
This kernel may be viewed as the wave function for an off-shell
winding state with ``momenta'' $\alpha^+$  and $\tilde\alpha^-$.
Equivalently, one may diagonalize the complementary set of operators,
\begin{equation} \alpha^+=-i\nu\partial_{\alpha^-},\ \ \
\tilde\alpha^-=-i\nu\partial_{\tilde\alpha^+} \label{oscrout}
\end{equation}
leading to on-shell wave  functions
\begin{equation} \label{eigoout} f( \alpha^-,  \tilde\alpha^+ ) = N_{out}
\left( \epsilon ~ \alpha^- \right)^{-\frac{M^2}{2i\nu} -
\frac12} \left( \tilde\epsilon~ \tilde\alpha^+
\right)^{- \frac{ \tilde M^2}{2i\nu} - \frac12} \end{equation}
Those are related  to the real-space representation  by the
kernel
\begin{equation} \label{phiout} \Phi^{out}_{ \nu, \tilde\alpha^+,
\alpha^-}(x^+,x^-) = \exp\left( - \frac{i\nu x^+
x^-}{2} - i \tilde\alpha^+ x^- -i \alpha^- x^+ -
\frac{i}{\nu} \tilde\alpha^+ \alpha^- \right)
\end{equation}
Replacing $f_1^*(x^+,x^-)$ and $f_3(x^+,x^-)$ by their expression 
in terms of the $out$ and $in$ 
wave functions \eqref{eigoin}, \eqref{eigoout} respectively,
renders the $x^\pm$ Gaussian (albeit with a non-positive definite
quadratic form). The remaining $\alpha^\pm,\tilde\alpha^\pm$ integrals
can now be computed in terms of hypergeometric functions. Including
the form factor from the excited modes, we obtain, for the 3-point 
amplitude,
\begin{multline}
\langle 1 | :e^{i (p_2^+ X^- + p_2^- X^+) }:   | 3  \rangle =
\frac{g_s}{2\nu} 
\, \delta_{\sum j_i}\, \delta\left(\sum p_i^\perp \right) ~
\exp\left[
- p_2^+ p_2^- \tilde \Delta(\nu)  \right]\\
\left(-p_2^+\right)^{\mu-1} \left(-p_2^-\right)^{\tilde \mu-1}
U\left(\lambda, \mu, i \frac{p_2^+ p_2^-}{\nu}\right)
U\left(\tilde \lambda, \tilde \mu,i \frac{p_2^+ p_2^-}{\nu}\right)
\label{a3}
\end{multline}
where the  non-locality parameter $\tilde \Delta(\nu)$ includes
the contribution of the quasi-zero-mode,
\begin{equation}
\tilde \Delta(\nu) = \psi(i\nu)+\psi(1-i\nu)-2\psi(1)
\end{equation}
The parameters of the Tricomi confluent hypergeometric functions $U$
appearing in \eqref{a3} are given by
\begin{align}
\label{lm}
\lambda &= \frac12+\frac{M_3^2}{2i\nu} &
\tilde\lambda &= \frac12+\frac{\tilde M_3^2}{2i\nu} \notag \\
\mu &= 1+\frac{M_3^2-M_1^2}{2i\nu} &
\tilde\mu &= 1+i\frac{\tilde M_3^2-\tilde M_1^2}{2i\nu}\ .
\end{align}
The amplitude is finite, and it is proportional
to the overlap of the zero-mode wave-functions, up to the smearing due to the 
form factor of the untwisted string in the background of the twisted string.
Similar expressions can be obtained for 3-point functions in superstring
theory involving an untwisted massless state.

\subsubsection*{Four-point amplitudes.}
The same techniques allow to compute 4-point amplitudes, which now
include an integral over the location of the 4-th vertex, as well as
on the relative boost parameter $v$ between the two untwisted vertices. 
The complete expression can be found in \cite{Berkooz:2004yy} and is somewhat abstruse,
however it is useful to consider the factorization limit $z\to 0$ where
$T(3),T(4)$ (resp. $T(1),T(2)$) come together:
\begin{multline}
\label{a40}
 \langle 1 |  T(2)  T(3) | 4 \rangle  \to g_s^2~
\delta_{-j_1 +j_2+j_3+j_4} ~\delta\left(-\vec p_1 +  \sum_{i=1}^3
\vec p_i\right)~\\
\int_{-\infty}^{\infty} \! dv ~e^{i(j_3-j_1)v} ~
~ \int dz d\bar z ~
\lvert z\rvert^{2\vec p_3   \cdot\vec p_4 +
\vec p_3 \cdot \vec p_3 -2}
\exp\left[ - \left( p_2^+ p_2^-+p_3^+ p_3^- \right)
\tilde \Delta(\nu)  \right] \\
(-1)^{\mu+\tilde\mu}~
(p_2^+)^{-\tilde\lambda}~
(p_2^-)^{-\lambda}~
(p_3^+)^{\mu-\lambda-1}~
(p_3^-)^{\tilde\mu-\tilde\lambda-1}~
z^{-\frac12 M_1^2 - \frac{i\nu}{2}}
\bar z^{-\frac12 \tilde M_1^2 - \frac{i\nu}{2}}
\end{multline}
The amplitude diverges whenever $j_3=j_1$
due to the propagation of winding strings
with vanishing boost momentum in the intermediate channel.
This result closely parallels the discussion in
Ref.~\cite{Berkooz:2002je}, where tree-level scattering amplitudes
of four untwisted states where found to diverge, due to large graviton
exchange near the singularity.

\subsection{More than two twisted strings}

Scattering amplitudes involving three  or more twisted states
can be obtained by mapping to an analogous problem which is 
now very well understood: the Wess-Zumino-Witten
model of a four-dimensional Neveu-Schwarz plane wave 
\cite{Nappi:1993ie,Olive:1993hk,Kiritsis:jk}, with metric
\begin{equation}
\label{ds}
ds^2  = -2 du dv + d\zeta d\tilde \zeta - \frac14 \zeta\tilde\zeta du^2\  ,\qquad
H= du dx  d\bar x
\end{equation}
where $\zeta=x_1+ix_2$  is the complex coordinate in the plane.
In the light-cone gauge $u=p \tau$, it is well known that the transverse
coordinate $X$ has the mode expansion of a complex scalar field twisted by  a
real, non rational angle proportional to the light-cone
momentum $p$~\cite{Kiritsis:jk}. In fact,
there exists a free-field representation where the vertex operator
of a physical state with non-zero $p$ is just the  product
of  a plane wave along the $(u,v)$ light cone coordinates,
times a twist field\footnote{For integer $p$, new ``spectrally flowed''
states appear describing long strings \cite{Kiritsis:2002kz}.}
creating a cut $z^p$ on the world-sheet.
Correlation functions of physical states
have been computed using standard WZW techniques
\cite{D'Appollonio:2003dr,Cheung:2003ym},
and, by removing the plane wave contribution, it is then possible
to extract the correlator of twist fields with arbitrary angle.

Referring the reader to \cite{Berkooz:2004yy} for more details, we simply quote
the result for the three twist amplitude: in real-space representation
\eqref{reals}, the amplitude (hence, the OPE coefficient of 3 twist fields)
is given by the overlap
\begin{eqnarray}
\label{ker3}
\int dx^\pm_1  dx^\pm_2~ &&
~\exp\left[ (x_1^+ - x_2^+)(x_1^- - x_2^-) \Xi(\nu_1,\nu_2) \right] 
\nonumber\\
&&   [ f_1(x_1^\pm) f_2(x_2^\pm) ]^* 
~f_3\left( x_3^\pm - \frac{\nu_1 x_1^\pm + \nu_2 x_2^\pm}{\nu_1+\nu_2}
\right)
\end{eqnarray}
where the characteristic size of the kernel is given by the ratio
\begin{equation}
\label{xi}
\Xi(\nu_1,\nu_2) = -i \frac{
1- \frac{i \nu_3}{\nu_1\nu_2}
\frac{ \gamma(i \nu_3)}{ \gamma(i \nu_1) \gamma(i \nu_2)}}
{1+ \frac{i \nu_3}{\nu_1\nu_2}
\frac{ \gamma(i \nu_3)}{ \gamma(i \nu_1) \gamma(i \nu_2)}}
\end{equation}
with $\gamma(p)\equiv \Gamma(p)/\Gamma(1-p)$.
As $\nu_i\to 0$, $\Xi(\nu_1,\nu_2) \sim 1/ (2\zeta(3)~\nu_3^2)$
so that the interaction becomes local, as expected for flat space
vertex operators. For larger $\nu$ however,
the non-locality scale $1/\sqrt\Xi$ diverges when
$\nu_1 \nu_2 \gamma(i \nu_1) \gamma(i \nu_2) =
i\nu_3 \gamma(i \nu_3)$. The origin of this divergence is not
well understood at present.

\subsection{Toward classical back-reaction}
While computing the back-reaction from the quantum production of
particles and strings remains untractable with the present techniques,
the results above give us a handle on a
related problem, namely the linear response of closed string fields to a 
classical (coherent) condensate of winding strings. Indeed, consider
deforming Misner space away from the orbifold point, by adding to the
free worldsheet action a condensate of marginal twist operators:
\begin{equation}
S_\lambda = \int d^2\sigma~ \partial X^+ \bar \partial X^- 
+ \lambda_{-w} V_{+w} + \lambda_{+w} V_{-w} 
\end{equation}
While this deformation is marginal at leading order, it implies
a one-point function for untwisted fields
\begin{equation}
\label{3p1}
\langle e^{ikX} \rangle_\lambda \sim \lambda_{w} \lambda_{-w}
\langle w |  e^{ikX} | -w \rangle\ ,\quad
\end{equation}
which needs to be cancelled  by deformating $S$ at order $\lambda^2$
by an untwisted field: this is the untwisted field classically 
sourced by  the winding string with vertex operator $V_{\pm w}$.
In addition, the same winding string also sources twisted states
whose winding number is a multiple of $w$:
\begin{equation}
\label{3p2}
\langle V_{-2w} \rangle_\lambda \sim \lambda_{w} \lambda_{w}
\langle w |  V_{-2w} | w \rangle\ ,\quad
\end{equation}
The 3-point functions in \eqref{3p1}, \eqref{3p2} are precisely the amplitudes
which have been computed the two previous sections. It is thus possible
to extract the corrections to the metric and other string fields to 
leading order in the deformation parameter $\lambda_{w}$. In the Euclidean
orbifold case, such a procedure allows to resolve a conical ALE singularity
into a smooth Eguchi Hanson gravitational instanton. 
Whether the same procedure allows to resolve the divergences of 
the Lorentzian orbifold remains an intriguing open question.

\section{Discussion}

In this lecture, we have taken a tour of the classical aspects of the
propagation of closed strings in a toy model of a cosmological singularity:
Misner space, a.k.a. the Lorentzian orbifold $\mathbb{R}^{1,1}/\mathbb{Z}$. Our
emphasis has been particularly on twisted sectors, which play such
an important r\^ole in resolving the conical singularities of Euclidean
orbifolds. In particular, we have obtained a semi-classical understanding
of the pair production of  winding strings, as a tunneling effect in the
Rindler regions, in close analogy to Schwinger pair creation in an electric
field. Despite the fact that the one-loop amplitude remains real, indicating
no overall particle production between infinite past and infinite future, it
is clear that abundant production of particles and strings takes place near
the singularity.

While tree-level scattering amplitudes exhibit severe divergences due
to the infinite blue-shift near the singularity, it is quite conceivable
that the back-reaction from the cosmological production of particles and 
winding strings may lead to a smooth cosmology, interpolating between
the collapsing and expanding phases. Indeed, winding strings behave much
like a two-dimensional positive cosmological constant, and may thus lead
to a transient inflation preventing the singularity to occur. 

Unfortunately, incorporating back-reaction from 
quantum production lies outside the
scope of current perturbative string technology at present. A second
quantized definition of string theory would seem to be a prerequisite
to even formulate this question, however, unlike the open string
case, a field theory of off-shell closed strings has remained elusive, 
and may even be excluded on general grounds. A generalization of the
usual first quantized approach allowing for non-local deformations of
the worldsheet \cite{Aharony:2001pa}
may in principle incorporate emission of correlated
pairs of particles, however do not seem very tractable at 
present. 

Instead, the most practical approach seems to consider classical 
deformations by twisted fields away from the orbifold point. 
In contrast to the problem of quantum back-reaction, this may be 
treated in conformal perturbation theory, and we have taken some steps 
in this direction. It remains to see whether Misner space is a
good approximation to the resulting space. 

More importantly, Misner space appears to be a very finely tuned example of the
space-like singularities which are generically expected to occur in classical
Einstein gravity: as shown long ago by Belinsky, Khalatnikov and Lifshitz,
and independently by Misner himself (see e.g. \cite{Damour:2002et} for a 
recent review), the generic approach to a cosmological
singularity consists of a chaotic sequence of ``Kasner'' epochs (of which Milne/Misner
space is a special example with zero curvature) and curvature-induced bounces,
occuring heterogeneously through space. An outstanding question is therefore
to understand string theory in Misner (Mixmaster) space.

\begin{acknowledgments}
Both authors are grateful to M. Berkooz, D. Reichman and M. Rozali
for a very enjoyable collaboration, and to B. Craps, G. d'Appolonio, 
E. Kiritsis, and G. Moore for useful discussions. 
B. D. would like to thank the organizers of the Carg\`ese 2004 ASI 
for giving him the opportunity to present
 part of this work in the Gong Show. 

\end{acknowledgments}

\numberwithin{equation}{section}
%\numberwithin{figure}{section}

\end{document}